\documentclass[12pt,letterpaper]{article}
\newcommand{\be}{\begin{equation}}
\newcommand{\ee}{\end{equation}}
\newcommand{\beq}{\begin{equation}}
\newcommand{\eeq}{\end{equation}}
\newcommand{\bea}{\begin{eqnarray}}
\newcommand{\eea}{\end{eqnarray}}
\newcommand{\beqa}{\begin{eqnarray}}
\newcommand{\eeqa}{\end{eqnarray}}

\newcommand{\ap}{\ensuremath{\alpha'}} 

\def\slash#1{#1\!\!\!\!/\ }
\def\mattrix#1{\ensuremath{\mathbf{#1}}}

\newcommand{\cM}{{\mathcal{M}}}

\newcommand{\cC}{{\mathcal{C}}}
\newcommand{\cH}{{\mathcal{H}}}
\newcommand{\cL}{{\mathcal{L}}}

\newcommand{\hodge}{\ensuremath{\hat{*}}}

\newcommand{\del}{\delta}

\newcommand{\s}{\sigma}

\newcommand{\eg}{{\it e.g.}}
\newcommand{\ie}{{\it i.e.}}

\newcommand{\reef}[1]{(\ref{#1})}

\newcommand{\non}{\nonumber}
\newcommand{\pf}{\partial}

\newcommand{\df}{\textrm{d}}

\begin{document}

\begin{titlepage}

\rightline{\small hep-th/0204103 \hfill UTPT-02-04} \vspace*{2cm}

\begin{center}
{\Large\bf Aspects of supertubes}

\end{center}
\vspace{2mm}

\begin{center}

{\large
Mart\'\i n Kruczenski,\footnote[1]{e-mail: \tt martink@physics.utoronto.ca}
Robert C. Myers,\footnote[2]{e-mail: \tt rcm@hep.physics.mcgill.ca}\\
\vspace{2mm}
Amanda W. Peet\footnote[3]{e-mail: \tt peet@physics.utoronto.ca} and
David J. Winters\footnote[4]{e-mail: \tt winters@hep.physics.mcgill.ca}}

\vspace{3mm}
 $^{1,3}$  Department of Physics, University of Toronto\\
       60 St. George St., Toronto, Ontario M5S 1A7, Canada \\
\vspace{3mm}
 $^{1,2}$  Perimeter Institute for Theoretical Physics \\
       35 King St. N., Waterloo, Ontario N2J 2W9, Canada \\
\vspace{3mm}
 $^2$  Department of Physics, University of Waterloo\\
       Waterloo, Ontario N2L 3G1, Canada \\
\vspace{3mm}
 $^{2,4}$  Department of Physics, McGill University\\
       Montr\'eal, Qu\'ebec H3A 2T8, Canada

\end{center}

\vspace{3mm}

\begin{center}
{\large \bf Abstract}
\end{center}
\noindent
We find supersymmetric solutions of the D4-brane Born-Infeld action
describing D2 supertubes ending on an arbitrary curve inside a D4-brane.
{}From the D4-brane point of view, these are dyonic strings.
We also consider various higher dimensional extensions of the
usual supertubes, involving expanded D4- and D3-brane configurations.
Finally, considering the worldsheet theory for open strings on a
supertube, we show that this configuration is an exact solution
to all orders in $\ap$. Further the causal structure of
the open-string metric provides new insight into the arbitrary
cross-section of the supertube solutions. {}From this point of view,
it is similar to the arbitrary profile that appears for certain null
plane waves.

\vspace{2mm}

\vfill
\begin{flushleft}
April 2002
\end{flushleft}
\end{titlepage}

\vspace{3mm}

\newpage

\section{Introduction}

Brane expansion is an interesting aspect of the physics of D-branes
which has been found to occur in a variety of contexts. In most cases,
this expansion is a dynamical effect that arises through the interaction
of the D-branes with external supergravity fields \cite{Myers:1999ps}.
Various examples of this effect include the polarization of
D-branes in Wess-Zumino-Witten backgrounds \cite{wzw}, giant gravitons
\cite{giant} and the supergravity resolution of singularities in
N=1$^*$ super-Yang-Mills theory \cite{matt}. A similar expansion
was found to be possible by stabilising an ellipsoidal D0-D2 bound
state with angular momentum \cite{Harmark:2000na}. This construction
provided the first example where the expansion was not due to
external fields. However, these configurations were not supersymmetric.

More recently, however, Mateos and Townsend \cite{Mateos:2001qs} constructed supersymmetric
configurations in which a D2-brane has expanded into a cylinder
in flat, empty space. Again, in contrast to the original examples, these
supertubes are supported against collapse solely by the excitation of internal
fields in the D2-brane worldvolume theory. More precisely, the (static)
electric and magnetic fields on the brane produce an angular momentum
which stabilises the cylinder at a finite radius.
Remarkably the cross-section of the cylinder can be an arbitrary curve
embedded in eight-dimensional space transverse to the axis of the
cylinder \cite{Mateos:2001pi,Bak:2001xx,curve}. That is, given worldvolume
coordinates $t$,$x$,$\phi$, the configuration
\bea
X_0 &=& t ,\ \ X_9=x, \ \  X_i = X_i(\phi) ,\ i=1\ldots 8
\nonumber\\
F &=& dt\wedge dx + B(\phi) dx\wedge d\phi ,
\label{eq:supertube}
\eea
where $X_i(\phi)$ and $B(\phi)>0$ are arbitrary functions, constitutes
a solution of the full Born-Infeld equations of the D2-brane worldvolume.
Further, all of these solutions still preserve $1/4$ of the type IIA
supersymmetries.
Here $F$ denotes the field strength of the worldvolume gauge field and it
is crucial that the electric field is one in string units ($F_{tx}=1$).
These electric and magnetic fields can be interpreted as indicating that
fundamental strings and D0-branes (respectively) have been `dissolved'
into the D2 worldvolume.
Further investigations revealed interesting T-dual configurations
corresponding to helical D-strings~\cite{Cho:2001ys}. Other
aspects of the physics of supertubes can be found in
\cite{Bak:2001tt,Bak:2001gm,Bena:2001wp}

Another related facet of D-brane physics is the possibility that
a D$p$-brane can, with the excitation of certain worldvolume
fields, `morph' into a D$p'$-brane, of lower or higher dimensionality.
For example, from Matrix theory \cite{matrix} or the dielectric effect
\cite{Myers:1999ps}, a collection of D0-branes can expand in to various
higher dimensional branes through non-commutative geometry.
Similarly, the above supertube configurations \reef{eq:supertube} can be
described as a non-commutative geometry within the worldvolume theory
of the constituent D0-branes \cite{Bak:2001xx,matrix2}.

Another example of D-brane `morphing' is that D-strings ending on a
D3-brane can be described with
remarkable accuracy, from the point of the worldvolume theory of the
D3-brane \cite{Callan:1997kz,others}, as a spike adorned with an
appropriate magnetic field. Remarkably, a complementary description
of this geometry as a `fuzzy funnel' \cite{fun} is also provided by
the worldvolume theory of the D-strings.
Now we may consider supertubes, composed of D2-branes, fundamental
strings and D0-branes.
Either of the first two constituents can end on D4-branes
in a supersymmetric way \cite{end}. Similarly, the supersymmetry of D0-
and D4-branes is compatible (see, \eg, \cite{bran}) and the D0's can
dissolve in the D4 worldvolume \cite{douglas}.
This suggests that a supertube can end on a D4-brane while preserving
1/8 of the supersymmetry. We will show this result in fact holds
with an explicit construction.

In the present paper, we examine various aspects of the physics of
supertubes. First, in section~\ref{sec:D2onD4}, we illustrate that
supertubes can end on D4-branes by constructing the appropriate
solution of the D4-brane worldvolume action, following
\cite{Callan:1997kz,others}. We verify that these configurations
solve the full nonlinear Born-Infeld equations and that they
preserve the expected 1/8 of the supersymmetries. The next two
sections describe attempts to produce nontrivial higher
dimensional extensions of the supertube. T-duality easily allows
one to construct supertubes which are expanded D$p$-branes where
the spatial geometry is a flat cylinder of the form $S^1\times
R^{p-1}$. In section~\ref{sec:D4st}, we show that in the D4-brane
case this geometry can be deformed such that the worldvolume
metric is no longer flat. In the following section, we construct
new D3-brane solutions where the spatial geometry is  $S^1\times
S^1\times R^1$, where the two orthogonal circles are each
supported by independent angular momenta. Unfortunately the latter
configurations are not supersymmetric. In section 5, we examine
the world sheet theory of open strings ending on a supertube and
show that it is conformal to all orders in $\ap$. We discuss how
the peculiar structure of the open-string metric
\cite{Seiberg:1999vs} provides new insight into the arbitrary
shape and magnetic field profile of the supertube
\reef{eq:supertube}. {}From this point of view, this result is
similar to the arbitrariness in choosing the profile of plane wave
excitation of the transverse scalars propagating on a D$p$-brane.
There is also a close analogy to the arbitrary profile appearing
in certain exact closed string backgrounds representing
gravitational waves \cite{Amati:1989}. Finally we give a brief
discussion of our results in section~\ref{sec:conclusions}.


\section{D2 supertubes ending on D4-branes.}
\label{sec:D2onD4}

As discussed above, one should expect that supertubes can end on an
orthogonal D4-brane while preserving 1/8 of the supersymmetry.
We verify this intuition with an explicit construction of such a
configuration within the worldvolume theory of the D4-brane. At leading
order, the low energy theory on a single D4-brane reduces to ordinary
Maxwell theory coupled to a set of massless scalars describing the
transverse position of the brane --- see, \eg, \cite{bran}. We begin
by constructing an appropriate solution of this leading order
theory. In the next subsection, we show that this configuration in
fact solves the full Born-Infeld equations of motion. Subsequently
we will also verify that the solution also preserves the expected
supersymmetries, also at the non-linear level ($\kappa$-symmetry).

So let us consider a curve $\cC$ inside
a flat D4-brane extending in the directions $(X_1,X_2,X_3,X_4)$ and find the
gauge field configuration that describes a supertube extending along $X_9$
and ending on $\cC$. In the D4 worldvolume, we use coordinates
$x_a, a=0,1,\ldots,4$. Space-like indices are denoted as
$i,j,\ldots=1,\ldots,4$.

As described in the introduction, the supertube has three types of
constituent branes: D2- and D0-branes, and fundamental strings. Our
construction will have to incorporate all of these components
by exciting the appropriate electromagnetic fields, as well as a
displacement of the worldvolume which is accomplished by exciting
the transverse scalar $X_9(x_a)$. Following \cite{Callan:1997kz},
the displacement and the fundamental strings can be described
by the following supersymmetric configuration
\be
F_{0i} = -\pf_i A_0 = \partial_i X_9(\vec{x}) , \ \ \
\Delta X_9(\vec{x}) = -4\pi^2 \rho(\vec{x}) ,
\label{eq:CM}
\ee
where the charge density $ \rho(\vec{x})>0$ has support on $\cC$ and
corresponds to the local density of strings ending on $\cC$.
The D2-branes will be represented by a (static) magnetic field which has the
curve $\cC$ as a monopole source:
\be
F' = \hodge dA , \ \ d\hodge dA = -4\pi^2\, \hodge j  ,
\label{eq:F1}
\ee
with $j$ a conserved current tangential to, and with support on, the curve $\cC$.
The symbol $\hodge$ denotes the Hodge dual in the four-spatial dimensions
using flat metric $\delta_{ij}$.
Since $j$ is conserved, $|j|$ is a constant corresponding to the
number of D2-branes ending on $\cC$, as follows from the
Chern-Simons coupling to the corresponding RR field strength.
If we look at the linearized supersymmetry conditions for the magnetic field
\be
\delta \chi = F_{ij} \Gamma^{ij} = 0   ,
\label{eq:linsusy}
\ee
we find out that this configuration is not supersymmetric but becomes so
if we add new components such that the magnetic field is self-dual,
\ie, $F=\hodge F$.
Therefore, we add to $F'$ a dual magnetic field:
\be
F = dA + \hodge dA ,\ \ d\hodge dA = -4\pi^2\, \hodge j  \ .
\label{eq:F2}
\ee
One easily verifies that the linearized equations of motion (\ie,
$d*F=0=dF$) are satisfied everywhere away from $\cC$. Notice
that $j$ can now also be considered as an electric current sourcing
the $dA$ component of the magnetic field.
The self-duality of this magnetic field also implies that $F\wedge F \neq 0$
which corresponds to the desired appearance of a density of
D0-branes.\footnote{Certain related solutions describing bound states
of D0-branes and fundamental strings stretching between two D4-branes
were described in \cite{tong}.}

The solution can be written explicitly as:
\bea
-A_0=X_9(\vec{x}) &=& \int \frac{\rho(\vec{x}')}{|\vec{x}-\vec{x}'|^2}
\,d^4x'\ ,
\nonumber \\
A_i(\vec{x})&=& \int \frac{j_i(\vec{x}')}{|\vec{x}-\vec{x}'|^2} \, d^4x'\ ,
\label{eq:solution}
\eea
which satisfies also the gauge condition $\partial_a A^a=0$.
Notice that the self-duality condition for the spatial part of $F$, and
the fact that $A_0=-X_9$, are required by supersymmetry but not by
the linearized equations of motion. Below we will find that these
conditions also play an important role in simplifying the full non-linear
Born-Infeld equations.

\subsection{Born-Infeld equations}

The Born-Infeld Lagrangian which controls the
low energy dynamics of the D4-brane worldvolume theory
may be written as:
\beq\label{lagrange}
\cL=-\tau_4\sqrt{-|g+F|},
\eeq
where $\tau_4 = \frac{1}{(2\pi)^{3/2}g_s}$ is the D4-brane tension
--- we have introduced units where the fundamental string tension is
unity, \ie, $2\pi\ap=1$. The induced metric on the worldvolume
is $g_{ab}=\eta_{\mu\nu}\pf_a X^\mu\pf_b X^\nu$
and $|\cdots|$ is used to denote the determinant of the enclosed
matrix. The full nonlinear
equations of motion of the worldvolume fields are then:
\bea
\partial_a {\cal M}^{[ab]} &=& 0
\nonumber\\
\partial_a \left({\cal M}^{(ab)} \partial_b X^{\mu}\right) &=& 0 ,
\label{eq:BIeq}
\eea
where
\be
{\cal M}^{ab} = \sqrt{|g+F|} \left((g+F)^{-1}\right)_{ab}  .
\label{eq:Mab}
\ee
For a generic static configuration, it is useful to write $g+F$ as:
\be
g+F =
\left(
\begin{array}{cc}
-1 & E^t \\
-E & M
\end{array} \right) \ ,
\label{eq:gpF}
\ee
where we have introduced $E_i\equiv F_{0i}$.
The inverse of $g+F$ is given by:
\be
(g+F)^{-1} = \frac{1}{\Delta}
\left(
\begin{array}{cc}
-1 & E^tM^{-1} \\
-M^{-1} E & M^{-1} \Delta + M^{-1}E \otimes E^tM^{-1}
\end{array}
\right) ,
\label{eq:gpFinv}
\ee
where $\Delta = 1-E^tM^{-1}E$.
For the present case,  $M_{ij} = \delta_{ij} + E_i E_j +F_{ij}$
since $\pf_iX_9=E_i$ by eq.~\reef{eq:CM}.
A useful fact for inverting $M_{ij}$ is that a self dual field
$F_{ij}=\frac{1}{2}\epsilon_{ijkl}F_{kl}$ satisfies
$F_{ij}F_{jk} = -\frac{1}{4} F^2 \delta_{ik}$.
Hence one finds
\be
\left(\mattrix{1} + E\otimes E+\mattrix{F}\right)^{-1} =
\frac{1}{1+\frac{1}{4}F^2}
\left\{\mattrix{1}-\mattrix{F}- \frac{1}{1+E^2+\frac{1}{4}F^2}
(E-\bar{E})\otimes(E+\bar{E})\right\} .
\label{eq:M1}
\ee
Here $\mattrix{F}$ is notation for $F_{ij}$ as a $4\times 4$  matrix and
$F^2 \equiv F_{ij}F_{ij}$. Also we introduced for convenience
$\bar{E}_i = F_{ij}E_j$, which
satisfies $\bar{E}E =0$, $\bar{E}\bar{E} = E^2 F^2/4$.

Finally a simple computation reveals that the determinant of $g+F$ is
independent of $E$ and takes the value:
\be
|g+F| = -\Delta\, |M| = -\frac{1+\frac{F^2}{4}}{1+E^2+\frac{F^2}{4}}(1+\frac{F^2}{4})(1+E^2+\frac{F^2}{4})
= - (1+\frac{1}{4}F^2)^2  .
\label{eq:det}
\ee
 Putting everything together, the final result for ${\cal M}$
has remarkably simple form:
\bea
{\cal M}^{00} &=& -(1+E^2+\frac{1}{4}F^2)
\nonumber\\
{\cal M}^{0i} &=&  E_i + \bar{E}_i
\nonumber\\
{\cal M}^{i0} &=& -E_i + \bar{E}_i
\nonumber\\
{\cal M}^{ij} &=& \delta_{ij} - F_{ij}  \ .
\label{eq:Mres}
\eea
Note that supersymmetry conditions from the previous section,
$E_i = \partial_i X_9$ and $F_{ij} = \frac{1}{2}\epsilon_{ijkl} F_{kl}$,
were essential ingredients in producing this simple form.
The equations $\partial_i {\cal M}^{[ij]} =0$ are satisfied if
$\partial_i E_i =0$ and $\partial_i F_{ij} =0$, which match
the Maxwell equations appearing at lowest order. Hence we are
assured that these equations are satisfied by the solution
given in eq.~\reef{eq:solution}.
The equations   $\partial_i \left({\cal M}^{(ij)} \partial_j X^{\mu}\right) =0$
 give:
\bea
\mu=0 &:& \partial_i(E_jF_{ji}) = \partial_{i}\pf_j X_9 F_{ji} + E_j
\partial_i F_{ji} =0 \nonumber\\
\mu=j &:& \partial_i({\cal M}^{(ij)}) =\partial_i \delta_{ij} = 0
\nonumber\\
\mu=9 &:& \partial_i\left({\cal M}^{(ij)}\partial_j X_9\right) = \
\partial^2 X_9 =0 \ .
\label{eq:BIeq2}
\eea
These are automatically satisfied except the last one, which
corresponds to the leading order scalar equation, and so is satisfied by the
given solution. Hence
we conclude that our configuration \reef{eq:solution}
satisfies all of the full nonlinear equations coming from the
 Born-Infeld action.

\subsection{Hamiltonian}

 It is also useful to compute the energy density. In order to do so we first compute the
momentum conjugate to $A_i$, which is
\be
\Pi_i = \frac{\partial {\cal L}}{\partial E_i} .
\label{eq:Pi}
\ee
 Using the notation of eq.(\ref{eq:gpF}) we get that for such a generic static
configuration
\be
\Pi_i  =  \tau_4 \sqrt{\frac{\det M}{\Delta}}\left( M^{-1}\right)_{(ij)} E_j .
\label{eq:Pi2}
\ee
 Using the properties of $M$ that we described above, we find the simple
result that, for our case, $\Pi_i=\tau_4 E_i$. Note that with this
result, the momentum density circulating in the world volume is
given simply by $T_{0i}= \Pi_j F_{ji}=-\tau_4\bar{E}_i $. Near the
supertube, we can consider a coordinate $\sigma$ along the
supertube and a radial coordinate in the transverse space. Then,
from eq.(\ref{eq:solution}) we find an electric field $E_r$ and a
magnetic field $F_{\sigma r}$ meaning that, in the vicinity of the
supertube, there is a non-vanishing  $\bar{E}_\sigma$. This gives
a momentum density along the supertube as expected\footnote{We
thank D. Mateos for related correspondence.}.

The Hamiltonian can also be computed with the result:
\be
{\cal H} = \sqrt{\tau_4^2 \det M + \Pi^t \bar{M} \Pi }  ,
\label{eq:H}
\ee
where $\bar{M} = (M^{-1}_{(ij)})^{-1}$, i.e. we take $M$ invert it, symmetrize and then
invert back. The result in general is different from $M$ but can be computed with the same
methods as before giving:
\be
\bar{M} = (1+\frac{1}{4}F^2)\,\mattrix{1} + E\otimes E -\frac{1}{1+E^2} \bar{E}\otimes \bar{E} .
\label{eq:Mbar}
\ee
The determinant of $M$ follows from eqs. (\ref{eq:gpFinv}) and (\ref{eq:det}):
\be
\det M = (1+\frac{1}{4}F^2)(1+E^2+\frac{1}{4}F^2) .
\label{eq:detM}
\ee
 Replacing in eq.(\ref{eq:H})  gives:
\be
{\cal H} = \tau_4 + \tau_4\frac{1}{4} F^2 + \frac{1}{\tau_4}\Pi^2 .
\label{eq:Hres1}
\ee
Integrating ${\cal H}$ over the D4 worldvolume, the total energy has three
separate (divergent) contributions.
The first term above yields the energy due to the D4-brane tension.
The term $\Pi^2/\tau_4$ yields that from the fundamental
strings and the $\tau_4F^2/4$, the energy from the D0-branes.
Note that the appearance of the latter two contributions, but
not a separate contribution for the D2-branes, matches the results
found in analysing the energy of the supertubes
\cite{Mateos:2001qs,Mateos:2001pi}. The result that the total
energy of the present configuration is a simple sum of a D4-brane
contribution and a supertube contribution reflects the
fact that 1/8 of the supersymmetries are still preserved.
 The energy integral diverges close
to the supertube. In that region it is convenient to choose an affine parameter
$\sigma$ along the supertube and spherical coordinates $r,\theta,\phi$ in the
transverse space. Excluding  a small region of radius $\epsilon(\sigma)$
around $\cC$ and integrating by parts using the equations of motion
we obtain that the leading contribution for $r\rightarrow 0$ region is:
\be
H = \int r^2 dr\, d\Omega_2\,d\sigma \,\cH
\, =\, 4\pi^3\tau_4\int d\sigma \left(\frac{j^2}{\epsilon(\sigma)}
+\frac{\rho}{\epsilon(\sigma)}\right)
= 2\pi \tau_2|j| \int d\sigma \left( \frac{|j|}{\rho}+\frac{\rho}{|j|}\right) X_9  ,
\label{eq:Hres2}
\ee
where we introduced $X_9=\pi\rho/\epsilon$ in the vicinity of the
singularity. Note that with our choice of units, the standard
D2-brane tension is given by $\tau_2=2\pi\tau_4$. Hence, as expected,
we find that the divergence is proportional to the distance, \ie, $X_9$,
by which the spike extends above the D4 worldvolume.
Identifying $2\pi|j|$ with the number of D2-branes as before,
it follows that we may identify $|j|/\rho$ and $\rho/|j|$ with $B$ and $\Pi$
of the supertube analysis \cite{Mateos:2001qs}. Then the expected
relation $\Pi B=1$ satisfied on the supertube follows.
Note that we are using the normalization of \cite{Mateos:2001pi}, where an
affine parametrization of the supertube is also used. We see that the freedom
in choosing $\rho$ is converted into the freedom in choosing $B$.

\subsection{Examples}

The solution can be better understood by considering two
examples\footnote{We have been informed that these examples were
studied independently by David Mateos and Selena Ng.}. One is the
case where $\cC$ is a straight line and the density of strings
$\rho$ is arbitrary, and other the case where $\cC$ is a circle
and $\rho$ is constant (which is the original supertube
of~\cite{Mateos:2001qs}).

 For $\cC$ a straight line we can take a coordinate $x$ along $\cC$,
and spherical coordinates $(r,\theta,\phi)$ in the transverse space. The integrals in
eq.(\ref{eq:solution}) give:
\bea
A_0 &=& -X_9 = -\int_{-\infty}^{\infty} \frac{\rho(x'+x)}{x'^2+r^2} dx'
= \frac{1}{r}\int \frac{\rho(x+yr)}{1+y^2}dy
\nonumber\\
A_x &=& \frac{\pi j}{r} . \label{eq:linear} 
\eea 
{}From here we can
compute the field strength: \be F= dx^0\wedge dX_9 +\frac{\pi
j}{r^2} dx\wedge dr + \pi j \sin\theta d\theta\wedge d\phi .
\label{eq:linearF} \ee The component $F_{\theta\phi}$ (coming from
$\hodge dA$) corresponds to a D2-brane along $r,x$, and the fields
$F_{0r}=\partial_r X_9$ and $F_{xr}=\pi j/r^2$ to the electric and
magnetic field of the supertube, respectively.

 The induced metric turns out to be:
\be
ds^2 = -dx_0^2 + dx^2 + dr^2 +r^2 d\Omega_2^2 +dX_9^2 ,
\ee
where $d\Omega_2^2$ is the line element on the two-sphere parametrised
by $\theta$, $\phi$. It is clear
that for $r\rightarrow\infty$ we have just the D4-brane. For $r\rightarrow 0$
we obtain from (\ref{eq:linear}) that $X_9\simeq \pi\rho(x)/r$ and then
$dX_9\simeq -(\pi\rho(x)/r^2) dr$. In this region
it is useful to use $X_9\simeq\pi\rho/r$ as a coordinate instead of $r$,  and
rewrite the field strength and metric as:
\bea
F &\simeq& dx_0\wedge dX_9 + \frac{j}{\rho} dx\wedge dX_9 +
\pi j \sin \theta d\theta\wedge d\phi
\nonumber\\
ds^2 &\simeq& -dx_0^2 +dx^2 + dX_9^2 \ .
\label{eq:linearFds}
\eea
 We see that the configuration reduces to a supertube with unit electric
field and magnetic
field equal to $j/\rho(x)$, as we inferred above from the analysis of
the energy. The metric in the $\theta$, $\phi$
directions is singular but the $F_{\theta\phi}$ component of the field strength
makes the action $\sqrt{g+F}$ non-singular and equal to that of the supertube.

The integrals can be done easily for a circular supertube with uniform magnetic field
which is the original supertube of~\cite{Mateos:2001qs}. Let us consider then a circle of radius $R$
and introduce polar coordinates $(r,\phi)$ in the plane of the circle and ($\rho,\theta$)
for the two remaining D4-brane coordinates. The density $\rho(\vec{x})$  will be taken
to be uniform, $\rho(\vec{x})=\rho_0 \delta(r-R) \delta(\rho)/(2\pi\rho)$, with $\rho_0$ a constant.
The result is:
\bea
 X_9 &=& -A_0 = \frac{2\pi R \rho_0}{\xi}
 \nonumber\\
 A &=& \frac{\pi j}{R} \left(\frac{r^2+\rho^2+R^2}{\xi}-1\right)\, d\phi \  ,
\label{eq:circ}
\eea
 where we defined $\xi^2 = (R^2 + \rho^2 + r^2)^2-4r^2R^2$.
 Near the circle $\xi$ goes to $0$ and
the solution becomes
\be
X_9 = -A_0 \simeq \frac{2\pi R \rho_0}{\xi},\ \ \ A \simeq \frac{2\pi j R}{\xi} \, d\phi  ,
\label{eq:circlin}
\ee
which is the same as for the flat supertube we described before, after proper identification
of the coordinates, \ie, $\xi \simeq 2R\sqrt{\rho^2+(r-R)^2}$.

\subsection{Supersymmetry}
\label{susy}

To complete the identification of the worldvolume solution with the configuration of
a D2 supertube ending on a D4-brane we should check that they preserve the same
supersymmetries. For the D4 worldvolume theory, the supersymmetry condition can be written
as~\cite{Bergshoeff:1997kr}:
\be
\Gamma \epsilon = \epsilon.
\label{eq:Gamma}
\ee
In our case,
\be
\Gamma =  \frac{1}{\sqrt{|g+F|}} \left\{
  1 + \frac{1}{2} F^{ab} \gamma_{ab}
    + \frac{1}{8} F^{ab} F^{cd} \gamma_{abcd}\right\}
  \Gamma_{11} \gamma_{01234} ,
\label{eq:ourGamma}
\ee
 where we introduced the worldvolume $\gamma$-matrices defined in terms
of the space-time matrices $\Gamma_{\mu}$ as:
\bea
\gamma_0 &=& \Gamma_0
\nonumber\\
\gamma_i &=& \Gamma_i + E_{i} \Gamma_9 \ ,
\label{eq:gammas}
\eea
Space-like indices should be risen and lowered with the metric:
\be
g_{ij} = \delta_{ij} + E_{i} E_{j} .
\label{eq:gind}
\ee
Using the self-duality of $F_{ij}$ a straightforward calculation gives:
\bea
 F^{ab} \gamma_{ab} &=&
  -\frac{2}{1+E^2} \Gamma_0 \left(\slash{E} + E^2 \Gamma_9\right) +
  \frac{2}{1+E^2} E_i\bar{E}_j \Gamma_{ij}
  \nonumber\\
 F^{ab} F^{cd} \gamma_{abcd} &=&
  8  \Gamma_0 \slash{\bar{E}}\Gamma_{1234} + \frac{2F^2}{1+E^2}
  \left(1+\Gamma_9\slash{E}\right)\Gamma_{1234}
  \nonumber\\
 \gamma_{01234} &=&\left(1+\Gamma_9\slash{E}\right) \Gamma_{01234}  ,
\label{eq:Fs}
\eea
where we used the notation $\slash{E} = E_i\Gamma_i$ and introduced again $\bar{E}_i = F_{ij}E_j$.
This gives finally
\bea
\Gamma &=& \Gamma_{11} \Gamma_{01234} +
\nonumber\\
&& + \frac{1}{1+\frac{1}{4}F^2}\Gamma_0\left[
\left(\frac{1}{2}F_{ij}\Gamma_{ij} - \slash{E}\Gamma_9
+\slash{\bar{E}}\Gamma_9 + \slash{\bar{E}}\slash{E}\right)
\left(\Gamma_{1234}-1\right)\right.
\nonumber\\
 &&\left. +\left(\slash{E}\Gamma_0\Gamma_{1234} + \slash{\bar{E}}\Gamma_9 +
\slash{\bar{E}}\slash{E}\right)\left(1+\Gamma_0\Gamma_9\Gamma_{11}\right)
\right] .
\label{eq:ourGamma2}
\eea
Since $F_{ij}$ and $E_i$ are not constant the only solutions of
$\Gamma \epsilon = \epsilon$ are those satisfying the three projections:
\bea
\Gamma_{11} \Gamma_{01234} \epsilon &=& \epsilon
\nonumber\\
\Gamma_{0} \Gamma_{11} \epsilon &=& \epsilon \  \ \ \
                     (\mathrm{or}\ \Gamma_{1234}\epsilon = \epsilon)
\nonumber\\
\Gamma_{09}\Gamma_{11} \epsilon &=& -\epsilon  \ .
\label{eq:susys}
\eea
Hence we find that 1/8 of the supersymmetries are preserved and
they match precisely with those expected for the D4-brane and the supertube.
In particular, the latter two conditions match those of D0-branes
and fundamental strings stretched along the $X_9$ axis. As expected,
there is no separate projection which we might associate with the
constituent D2-branes of the supertube \cite{Mateos:2001qs}.

\section{D4 supertubes}
\label{sec:D4st}

In this and the next section, we consider higher dimensional configurations
which can be thought of as nontrivial extensions of the supertube.
A natural way to increase the dimension of the D-branes is to apply T-duality
in directions transverse to the original supertube configuration
\reef{eq:supertube}. Suppose that such supertube extends along $X_9$ and
the cross section ${\cal C}$ is embedded  in the directions ($X_1,X_2,X_3,
X_4$). Performing two T-dualities along $X_5$ and $X_6$,
we obtain a D4 supertube with the supersymmetries of the fundamental strings
along $X_9$ and that of D2-branes filling the $X_5$-$X_6$ plane.
For the D2 supertube we can choose the magnetic field and the shape,
which amounts to choosing a distribution of D0-branes.
However the moduli space of the D2-branes in our D4 supertube is larger
than that of the D0-branes. Not only can we choose their positions but also
their orientations as a function of $\phi$. For the resulting configuration
to be supersymmetric the D2-branes must have
a common supersymmetry. That will be the case if they are related by an $SU(2)$
rotation~\cite{Berkooz:1996km}. We will consider a specific example in some
detail to understand the procedure better.
However, as is shown below, this case is singular at infinity
since there the energy density diverges   but we can consider similar
examples where the D2-branes are wrapped on a compact cycle of some
internal manifold which makes this problem disappear.

Consider then the following embedding:
\bea
&& X_1 + i X_2 = R e^{i\phi}, \ \ X_5 + i X_6 = y_1 e^{i\phi},
\ \ X_7 +i X_8  = y_2 e^{i\phi}
\nonumber\\
&& X_0 = t,\  \  X_9 = x,\ \   X_{3,4} =  0,
\label{eq:D4sup}
\eea
 and the worldvolume gauge field:
\be
 F=dt\wedge dx + B(\phi) dx\wedge d\phi ,
\label{eq:D4supF}
\ee
with $B(\phi)>0$.
The induced metric is
\be
ds^2 = -dt^2 + dx^2 + (R^2+y_1^2 + y_2^2)\, d\phi^2 + dy_1^2 + dy_2^2 .
\label{eq:D4supds}
\ee
The only difference with the T-dual of the D2 supertube is that
$g_{\phi\phi}$ depends on the extra coordinates $y_{1,2}$. It interesting to
observe that, as opposed to the D2 supertube, the induced
metric is {\it not} flat. Nevertheless, it is easy to see that the BI equations
are still satisfied.
Computing again ${\cal M}^{ij} = \sqrt{|g+F|}((g+F)^{-1})_{ij}$ we obtain that the only non-vanishing
components are:
\bea
&&\cM^{00} =   -\frac{f+B^2}{B} ,\ \  \cM^{0x} = -\cM^{x0} = \frac{f}{B} ,
\ \  \cM^{0\phi} = \cM^{\phi 0} = -1
\nonumber\\
&&\cM^{xx} = \frac{f}{B} ,\ \  \cM^{x\phi} = -\cM^{\phi x} = -1,
\nonumber\\
&&\cM^{y_1y_1} = \cM^{y_2y_2} =  B(\phi)  ,
\label{eq:D4supM}
\eea
with $f=R^2+y_1^2+y_2^2$. The equations $\partial_i {\cal M}^{[ij]}=0$ are
satisfied since $f$ and $B$ are independent of $x$ and $t$. The equations
$\partial_i ({\cal M}^{(ij)} \partial_jX^{\mu})=0$ are also satisfied because
$B$ is independent of $y_{1,2}$. For example, the equations for $\mu=5,6$
reduce to:
\be
\partial_{\phi}\left({\cal M}^{\phi\phi} \partial_{\phi} X^{5,6}\right)
+\partial_{y_1}
\left( {\cal M}^{y_1y_1}\partial_{y_1}X^{5,6} \right)=0 .
\label{eq:D4supBIeq}
\ee
 The second term vanishes if $B$ and $\partial_{y_1}X^{5,6}$ are independent of
 $y_1$. This is satisfied
if $X^{5,6}$ is linear in $y_1$, which
(together with a similar condition for $X^{7,8}$)
implies that the D2-branes are flat. The first term is zero
 since ${\cal M}^{\phi\phi}=0$,
which is due to the fact that $F_{tx}=1$. This is crucial because it allows
$X^{5,6}$ to depend on $\phi$.

 We see that the D2-branes can in fact point in arbitrary directions. However
we need that all the D2-branes preserve some common supersymmetry, which is
true if they are related by an $SU(2)$ rotation~\cite{Berkooz:1996km}.
 The computation of $\Gamma$ is easier than in the previous section and gives:
\be
\Gamma = \frac{1}{B} \left(1+\left(\gamma^{0x}\Gamma_{11}
         + B\gamma^{x\phi}\Gamma_{11}\right)\right)\Gamma_{11}
     \gamma_{tx\phi y_1y_2} .
\label{eq:D4supGamma}
\ee
The condition $\Gamma\epsilon =\epsilon$ is satisfied if:
\bea
\Gamma_{0}\Gamma_{9}\Gamma_{11}\epsilon &=& \epsilon
\nonumber\\
\Gamma_0 \gamma_{y_1}\gamma_{y_2} \epsilon &=& \epsilon .
\label{eq:D4supsusycond}
\eea
The latter condition reduces to
\bea
\Gamma_0\Gamma_6\Gamma_8 \epsilon &=& \epsilon
\nonumber\\
\Gamma_0\Gamma_5\Gamma_7 \epsilon &=& \epsilon .
\label{eq:D4supsusys}
\eea
These conditions are equivalent to those for fundamental strings along $X_1$
and D2-branes along $X_{4,6}$ and $X_{5,7}$.
Hence this configuration (\ref{eq:D4sup},\ref{eq:D4supF}) preserves $1/8$ of
IIA supersymmetry.

If we compute the Hamiltonian we encounter a problem. Indeed the energy is
given by:
\be
H = \tau_4\int d\phi\, dx\, dy_1\,dy_2 \sqrt{R^2+\vec{y}^2} \left( \frac{B(\phi)}{\sqrt{R^2+\vec{y}^2}}
+ \frac{\sqrt{R^2+\vec{y}^2}}{B(\phi)} \right) .
\label{eq:D4supH}
\ee
We see from above that the energy density diverges as $|\vec{y}|\rightarrow\infty$. This is due to the
fact that as we get away from the centre of the helix, the density of D2-branes decreases and then
the D4-brane becomes critical ($\det{\sqrt{g+F}}\rightarrow 0$). In the T-dual picture this corresponds
to a brane that at infinity moves at the speed of light. Clearly this problem
is associated with the infinite extent of the D2-branes and so it may
be avoided by considering a compact configuration.
If we compactify the some of the directions on a torus, K3 or Calabi-Yau
manifold then part of the
supersymmetry is preserved and we can wrap the D2-brane along some supersymmetric cycle. Again, the
moduli of such cycle can vary as a function of $\phi$ as long as some common supersymmetry is preserved.
 If the two-cycle contains non-trivial $S^1$ cycles then the moduli space includes also Wilson lines.
 For example in the case of $T^4$ we can wrap the D2-brane in a genus $g$ surface whose moduli space is
$T^4\times \mathrm{Symm}^g(T^4)$~\cite{Vafa:1997pm}. This configuration can also be described as that of $g$
intersecting supertubes preserving $1/8$ of the supersymmetry. If we consider $K3$ then the moduli
space for a genus $g$ surface is given by $\mathrm{Symm^g(K3)}$
\cite{Vafa:1997pm,Bershadsky:1995qy}.
Again the moduli can vary as a function of $\phi$ giving a large number of supertubes constructions.
On the other hand, these moduli spaces can be considered as
being related to the position of a D0-brane in a T-dual picture~\cite{Bershadsky:1995qy}. It would
be interesting to see if in the case of six-dimensional manifolds there are examples related to rotations
as in flat space.


\section{D3-tubes}

Applying a single T-duality transverse to the original supertube,
we get a D3-brane with D1-branes  and fundamental
strings dissolved at right angles. Embedded in flat space,
the T-dual configuration will have spatial topology $S^1\times R^2$,
where the $S^1$ is supported by angular momentum. In the following,
we consider a configuration with topology $S^1\times S^1\times R$ where
the orthogonal circles are both supported by separate angular momenta.
However, we will find that while this solution is stable, it is not
supersymmetric.

Consider Minkowski space in the following coordinates
\beq\label{minkmet}
ds^2=-dT^2+dX^2+dR_1^2+R_1^2 d\Phi_1^2+dR_2^2+R_2^2 d\Phi_2^2+dE_5^2,
\eeq
where the $R_i$ and $\Phi_i$ are radial and angular coordinates on two mutually
orthogonal planes and $E_5$ denotes five dimensional Euclidean space. Our
D3-tube, so-called, will be a D3-brane with one extended
($x$) and two compact ($\phi_1$, $\phi_2$)
spatial worldvolume directions, along with time ($t$).
We embed it in Minkowski space, using static gauge to align the worldvolume
and background coordinates as follows: $t=T$, $x=X$, $\phi_i=\Phi_i$ and
we will allow the radii to vary in each plane $R_i=R_i(\phi_i)$.
The tube sits, point-like, at the origin of the transverse $E_5$.
The Born-Infeld action for this D3-brane will take the form
\beq\label{action}
S=-\int\df^3\s \sqrt{-\det |g+ F|}
\eeq
(where we set $\tau_3=1$ in the following for convenience). With the
embedding described above, the induced metric $g_{ab}$ on the
the worldvolume becomes:
\beq\label{inducmet}
g_{tt}=-1, \quad g_{xx}=1, \quad g_{\phi_1\phi_1}=R_1^2+{R_1'}^2,
\quad
g_{\phi_2\phi_2}=R_2^2+{R_2'}^2,
\eeq
where the primes denote differentiation with respect to the appropriate angular
coordinate, \ie, $R_i'=\pf_{\phi_i} R_i$.

To induce separate angular momenta on each circle, we consider the
worldvolume gauge field \beq\label{F} F=E\df t \wedge \df x +
B_1\df x \wedge \df \phi_1 +B_2\df x \wedge \df \phi_2. \eeq This
corresponds to switching on an axial electric field, $E$,  and
introducing a magnetic flux, proportional to $B_i$, across each of
the compact circles. We will assume that $E$ is constant across
the entire worldvolume and that each of the magnetic components
only varies around its associated loop, \ie, $B_i=B_i(\phi_i)$.
{}From a microscopic perspective, we may think of $E$ as arising
from a uniform density of dissolved fundamental strings running
parallel to the axis of the tube. The magnetic fields are
associated with dissolved D1-branes wrapping each of the
orthogonal circles. As the system contains orthogonal D1-branes,
one should expect that no supersymmetries are preserved. In fact,
this intuition can be verified with a detailed calculation.

Having established our ansatz, we now write out the Lagrangian density,
\beq\label{lag}
\cL=-\sqrt{(1-E^2)(R_1^2+{R_1'}^2)(R_2^2+{R_2'}^2)+B_1^2(R_2^2+{R_2'}^2)+
B_2^2 (R_1^2+{R_1'}^2)},
\eeq
and perform the standard analysis of the Euler-Lagrange equations,
the components of the gauge potential $A$, and the transverse
scalars corresponding to the radii $R_i$. Due to our assumptions
about the uniformity of the various fields, the equations of motion
for $A_t$, $A_{\phi_1}$ and
$A_{\phi_2}$ are automatically satisfied. The equation of motion
for the remaining component $A_x$ becomes:
\beq\label{axeom}
0=\pf_{\phi_1}\left\{\cL^{-1}B_1(R_2^2+{R_2'}^2)\right\}
   +\pf_{\phi_2}\left\{\cL^{-1}B_2(R_1^2+{R_1'}^2)\right\}\ .
\eeq For the scalar $R_1$, one finds: \beq\label{r1eom}
\cL^{-1}\left((1-E^2)R_1(R_2^2+{R_2'}^2)+R_1B_2^2\right)=
\pf_{\phi_1}\left\{\cL^{-1}\left((1-E^2)R_1'(R_2^2+{R_2'}^2)+R_1'B_2^2\right)
\right\}\ . 
\eeq 
{}From the obvious symmetry of the Lagrangian
density \reef{lag}, the equation of motion for $R_2$ follows from
that for $R_1$ by the interchange ($1\leftrightarrow 2$) in all
subscripts. Carrying out the derivatives and simplifying each
equation in turn leads to \beqa\label{simpeom}
&&\{(1-E^2)(R_2^2+{R_2'}^2)+B_2^2\}(R_2^2+{R_2'}^2)\left(B_1'(R_1^2+{R_1'}^2)
-B_1R_1'(R_1+R_1'')\right)
+\non\\
&&\quad\{(1-E^2)(R_1^2+{R_1'}^2)+B_1^2\}(R_1^2+{R_1'}^2)\left(B_2'
(R_2^2+{R_2'}^2)-B_2R_2'(R_2+R_2'')
\right)=0,\non\\
&&\{(1-E^2)(R_2^2+{R_2'}^2)+B_2^2\}\left(\cL^{-1}(R_1-R_1'')-R_1'
\pf_{\phi_1}\cL^{-1}\right)=0,\non\\
&&\{(1-E^2)(R_1^2+{R_1'}^2)+B_1^2\}\left(\cL^{-1}(R_2-R_2'')-R_2'
\pf_{\phi_2}\cL^{-1}\right)=0.
\eeqa
It is apparent that each equation is satisfied if we choose
\beq\label{oursol}
(1-E^2)(R_i^2+{R_i'}^2)+B_i^2=0 \qquad (i=1,2).
\eeq
However, this is presumably not the only solution.
We could choose to set the other
bracket in each $R_i$ equation to
zero. Then the $A_x$ equation reduces to
\beq\label{hardsol}
\frac{B_1'}{R_1^2+{R_1'}^2}+\frac{B_1(R_1-R_1'')}{R_1'(R_1^2+{R_1'}^2)}=
\frac{B_2'}{R_2^2+{R_2'}^2}+\frac{B_2(R_2-R_2'')}{R_2'(R_2^2+{R_2'}^2)}=C,
\eeq
where $C$ is a constant, because each of the other terms is a function of
independent variables. Hence, solving the differential equation
\beq\label{de}
\frac{B'}{R^2+{R'}^2}+\frac{B(R-R'')}{R'(R^2+{R'}^2)}-C=0
\eeq
will also provide a solution to all of the equations. However, as
we were unable to make further progress towards solving this equation,
we focus on the solutions given by eq.~\reef{oursol}.

This solution states that $B_i^2=(E^2-1)(R_i^2+{R_i'}^2)$, which requires
$E^2\ge 1$ and further implies that
\beq
\label{correlate}
\frac{B_1^2}{ R_1^2+{R_1'}^2}=\frac{B_2^2}{ R_2^2+{R_2'}^2}=E^2-1\ .
\eeq
We also have periodic boundary
conditions on the $B_i$, the
$R_i$, and their derivatives. Solutions are easily constructed by finding a
cross-section
($R_1(\phi_1$), $R_2(\phi_2)$) that satisfies the periodic boundary conditions
and that is associated,
via eq.~\reef{oursol}, with magnetic fields that also satisfy the boundary
conditions (this translates to a
condition on each $R_i''(\phi_i)$). Then the magnetic fields are determined
up to a common factor $E^2-1$, which is arbitrary up to being non-negative.

Given that these configurations are not supersymmetric, an interesting
question is to determine whether or not they are stable.
To analyse this point, we change to the Hamiltonian
formalism. The only nontrivial canonical
momentum in the problem is that associated with $A_x$:
\beq\label{mom}
\Pi=\frac{\pf\cL}{\pf(\pf_0 A_x)}=\frac{\pf\cL}{\pf E}=-E\cL^{-1}
(R_1^2+{R_1'}^2)(R_2^2+{R_2'}^2).
\eeq
We then write the electric field in terms of its canonical momentum, or
`electric displacement', as
\beq\label{efrompi}
E=\Pi\sqrt{\frac{f_1f_2+B_1^2f_2+B_2^2f_1}{f_1f_2[\Pi^2+f_1f_2]}},
\eeq
where, for the sake of brevity, we have defined $f_i\equiv R_i^2+{R_i'}^2$
for $i=1,2$. Subsequently, the Hamiltonian density is written as
\beq\label{hamdens}
\cH=\Pi E-\cL=\sqrt{\frac{[\Pi^2+f_1f_2][f_1f_2+B_1^2f_2+B_2^2f_1]}{f_1f_2}}.
\eeq
Hamilton's equations then tell us, among other things, that the
`on-shell' expression for $\Pi$ is
\beq\label{pieom}
\Pi=E\sqrt{\frac{f_1f_2}{E^2-1}},
\eeq
which can be verified by substituting eq.~\reef{oursol} into eq.~\reef{mom}.
Since all the fields are independent of $x$, we can integrate the Hamiltonian
density over a cross-section
to get an energy per unit length that is uniform along the tube:
\beq\label{nrgperlength}
H=\int\df\phi_1\df\phi_2\;\sqrt{f_1f_2}\cH=\int\df\phi_1\df\phi_2\;
\sqrt{[\Pi^2+f_1f_2][f_1f_2+B_1^2f_2+B_2^2f_1]}.
\eeq

Corresponding to each solution is a triple of conserved quantities ---
the fluxes of $B_1$, $B_2$ and $\Pi$ across the torus:
\beqa\label{charges}
N^D_i&=&\int\df\phi_i\sqrt{f_i} B_i,\qquad(i=1,2)\\
N^F&=&\int\df\phi_1\df\phi_2\sqrt{f_1f_2} \Pi.\non
\eeqa
That is, for a given configuration, the number (density) of D1-branes
wrapping each of the circles is fixed, as is the number (density)
of fundamental strings along the $x$ axis.

To test the stability of the solutions we should consider the
functional second derivatives of $H$ with respect to variations in
all the fields. However, physical fluctuations will be constrained
by the conservation of $N^D_i$ and $N^F$. {}From $\del N^F=\del
N^D_i=0$ it follows that \beqa\label{depend} \del
\Pi&=&\left(\frac{d}{d\phi_1}\frac{\Pi R'_1f_2}{\sqrt{f_1f_2}}
          -\frac{\Pi R_1f_2}{\sqrt{f_1f_2}}\right)
      \frac{\del R_1}{\sqrt{f_1f_2}}
          +\left(\frac{d}{d\phi_2}\frac{\Pi R'_2f_1}
          {\sqrt{f_1f_2}}-\frac{\Pi R_2f_1}{\sqrt{f_1f_2}}\right)
      \frac{\del R_2}{\sqrt{f_1f_2}},\non\\
\del B_i&=&\left(\frac{d}{d\phi_i}\frac{B_i R'_i}{\sqrt{f_i}}
          -\frac{B_i R_i}{\sqrt{f_i}}\right)\frac{\del R_i}{\sqrt{f_i}}.
\eeqa
Using these expressions, we regard $\Pi$ and $B_i$ as functionals of the radii when calculating the variations
in $H$. For now, we consider two cases only: general (flux-preserving) fluctuations of a D3-tube with uniform
radii, and uniform fluctuations of a D3-tube with a general cross-section. In the
first instance we have $R_i'=0$, and in the second we have $\del R_i'=0$; these two assumptions greatly simplify the
functional differentiation. Evaluated on the solution space of eq.~\reef{oursol}, the matrix of second derivatives
for each case can be written as
\beq\label{Hdiffeom}
\mattrix{H}_{ij}\equiv\frac{\del^2 H}{\del R_i \del R_j}
=\frac{8 R_1 R_2 \sqrt{E^2-1}}{2E^2-1}\left(\begin{array}{ccc} 2 E^2 (R_1 f_2) / (R_2 f_1) & & 1 \\
                                                               1 & & 2 E^2 (R_2 f_1) / (R_1 f_2)
                                                 \end{array} \right),
\eeq
which has an obvious simplification in the case of uniform radii. Given that $E^2\ge 1$, the eigenvalues of
$\mattrix{H}$ can be shown to be non-negative, implying stability of the system against small (flux-preserving)
fluctuations of the fields, despite the loss of supersymmetry.


\section{Open strings on supertubes}

As mentioned in the introduction, one can choose an arbitrary
function $B(\phi)$ and an arbitrary shape (as a function of
$\phi$) for the supertube and still have a supersymmetric solution
of the Born-Infeld equations. To gain further insight into this
curious aspect of supertubes, we investigate the worldsheet theory
of open strings ending on a supertube. {}From this point of view,
this arbitrariness means that, for example, for any boundary term
associated with a magnetic field $F_{X_1\phi}(\phi)$, \bea
S_{\mathrm{bdy.}}&=&\int_{\partial \Sigma} d\tau {\cal V}_{X_1\phi}(\tau) \\
{\cal V}_{X_1\phi}(\tau) &=& \int dk \tilde{A}_1(k) e^{ik\phi}
\partial_\tau X_1
\eea
${\cal V}_{X_1\phi}(\tau)$ must have conformal dimension one since in (\ref{eq:supertube}),
any field $B(\phi)$ satisfies the equations of motion. The only way
this can be is if $\phi$ is a `null' field by which we
mean the correlator $\langle \phi \phi \rangle$ vanishes.
This would ensure that there is no anomalous dimension associated
with $e^{ik\phi}$, and so it has conformal dimension zero for any $k$.
This is analogous to the usual statement that $e^{i k_\mu X^\mu}$ has
conformal dimension $0$ if $k^2=0$, \ie, $\vec{k}$ is null.
Certainly we expect that $\phi$ should be a standard worldsheet
field with a nonvanishing propagator $\langle \phi \phi \rangle\propto
g^{\phi\phi}$. However, while this intuition is appropriate in the bulk of
the worldsheet, the above discussion refers to an interaction introduced on
the boundary and so the relevant metric for the boundary correlator
is the open-string metric as defined in~\cite{Seiberg:1999vs}. The latter
is modified by the background electric and magnetic fields and we will
see below that it indeed produces the desired result
$\langle \phi \phi \rangle\propto G^{\phi\phi}=0$.

With this motivation in mind, we analyse the world sheet action of open strings ending
on a supertube. Furthermore, as a by-product, we also prove that the supertubes are solutions
to all orders in $\ap$. This means that the solution does not have $\alpha'$
corrections but can have string loop corrections. This last point follows from an
analysis similar to those of \cite{Amati:1989,Horowitz:1994rf,Thorlacius:1997zd}.

Our strategy is to consider first a flat supertube with constant magnetic field and then
deform it with expectation values for the magnetic field and transverse deformations.

Consider then a supertube extending along directions $X_{1,2}$ and with
a field strength:
\be
(2\pi\ap)\, F = dX_0\wedge dX_1 + B\, dX_1\wedge dX_2,
\label{eq:fsflat}
\ee
(We restore factors of inverse string tension $2\pi\ap$ and the string
coupling $g_s$ in this section.)
The worldsheet action is given by:
\be
S = \frac{1}{4\pi\ap} \int_{\Sigma} \eta_{\mu\nu} \partial X^\mu \bar{\partial} X^\nu -\frac{i}{2}
\int_{\partial\Sigma} F_{\mu\nu} X^{\mu} \partial_{\tau} X^{\nu}
\label{eq:wsac}
\ee
 This is a free theory with boundary propagators given by~\cite{Fradkin:1985,Callan:1987}:
\be
\langle X^{\mu}(\tau) X^{\nu}(\tau') \rangle
=  -\ap G^{\mu\nu} \ln(\tau-\tau')^2
+ \frac{i}{2} \theta^{\mu\nu} \epsilon(\tau-\tau') ,
\label{eq:prop}
\ee
where $\epsilon(\tau)=\mathrm{sign}(\tau)$. Also, $G^{\mu\nu}$ and $\theta^{\mu\nu}$ are a symmetric
and anti-symmetric matrix respectively and are given by:
\be
G^{\mu\nu} + \frac{1}{2\pi\ap} \theta^{\mu\nu} = \left(\frac{1}{\eta+2\pi\ap F}\right)^{\mu\nu}
\label{eq:opmetric}
\ee
The propagators for the transverse coordinates are not modified by the gauge field, so we will
concentrate on $\mu,\nu=0,1,2$. Using the value of $F$ given by eq.(\ref{eq:fsflat}) we obtain that
the non-vanishing components of $G^{\mu\nu}$ and $\theta^{\mu\nu}$
are:\footnote{Related expressions appeared in~\cite{Bak:2001xx}. }
\bea
G^{00} &=& -\frac{1+B^2}{B^2}, \ G^{02} = G^{20} = -\frac{1}{B}, \
G^{11} = \frac{1}{B^2},
\nonumber\\
\theta^{01} &=& -\theta^{10} = -\frac{2\pi\ap}{B^2}, \
\theta^{12}=-\theta^{21} = \frac{2\pi\ap}{B}.
\label{eq:opmetric2}
\eea
Notice that $G^{22}=0$ and hence the boundary correlator
$\langle X^2(\tau) X^2(\tau') \rangle $ vanishes, as desired.
We can also compute
the open-string coupling constant $G_0$ as~\cite{Seiberg:1999vs}:
\be
G_o = g_s \left(\det(\eta+2\pi\ap F)\over\det(\eta)\right)^{\frac{1}{2}} = g_s B.
\ee
Therefore, we can trust the open-string picture as long as $g_s B \ll 1 $ and $g_s\ll 1$ to suppress
closed string loops. If we consider $N$ branes then we need $g_s B N\ll 1$.

Going back to the open-string metric, at this point it is convenient to define new coordinates as:
\be
\tilde{X_0} = B X_0 - \frac{1+B^2}{2} X_2, \ \ \tilde{X}_1 = B X_1 .
\label{eq:tildecoord}
\ee
The open-string metric can be written now as
\be
dS^2 = G_{\mu\nu} dX^{\mu} dX^{\nu} = -2 d\tilde{X}_0 dX_2 + d\tilde{X}_1^2 .
\label{eq:opds}
\ee
Notice that $\tilde{X}_0$ and $X_2$ are both null coordinates in this metric.

Now we can consider deformations of the conformal theory corresponding to
the addition of a position-dependent magnetic field, as well as changing the
shape of the supertube. The worldsheet action should now
include terms
\be
S_I = i \int_{\partial\Sigma} A_1(X_2) \partial_\tau X_1 + i
\int_{\partial\Sigma} \Phi_i(X_2) \partial_{\sigma} X_i .
\label{eq:SI}
\ee
For small values of $A_1(X_2)$ and $\Phi_i(X_2)$ all that is needed is that the extra terms have conformal dimension
$1$ in the unperturbed theory. This follows from the fact that $\langle X^2(\tau) X^2(\tau') \rangle=0$
provided that there are no contractions between $X_2$ and $\partial_\tau X^{1}$. These contractions are
proportional to $\partial_\tau\epsilon(\tau-\tau')=\delta(\tau-\tau')$ and vanish in a point splitting
regularization~\cite{Seiberg:1999vs}.

We want, however, to go a step further and consider the whole perturbative series showing that the
$\beta$ functions $\beta^{A_1}$ , $\beta^{\Phi_i}$ are $0$ to all orders in $\ap$. The basic observation
is that $X_2$ is a null coordinate in the open-string metric (\ref{eq:opds}). This means that
the background is analogous to a plane wave and we can use the same methods.

 One way to show that the fields $A_1(X_2)$ and $\Phi_i(X_2)$ are not renormalized is to expand $S_I$
around a background field $X^\mu=\bar{X}^\mu+x^\mu$.  Then one can show that terms proportional
to $\partial_\tau \bar{X}^1$ or $\partial_\sigma \bar{X}^i$, which could renormalize
$A_1(X_2)$ or $\Phi_i(X_2)$, cannot be generated in one-particle-irreducible
vacuum Feynman diagrams since there is no
$\langle X^2 X^2\rangle$ propagator.

Another way is to proceed directly to compute the partition function with sources. As discussed
in~\cite{Seiberg:1999vs} the non-commutativity produces an overall factor in the computation
of vertex correlators and so the computations can be done in the commutative case introducing
the parameters $\theta^{\mu\nu}$ at the end.
Furthermore, we just need to  consider the boundary theory since in the bulk of the
world sheet there are no divergences.  The boundary theory with sources is simply
\bea
S &=& \int_{\partial\Sigma,\partial\Sigma'} d\tau\,d\tau'\, X^2_{\tau}
\,G^{-1}_{\tau,\tau'}\,\tilde{X}^0_{\tau'}
                           + \tilde{X}^1_{\tau}\,G^{-1}_{\tau,\tau'}\,
               \tilde{X}^1_{\tau'} +
    \partial_{\sigma} X^i_{\tau} \left(\partial_{\tau\tau'}G_{\tau,\tau'}
    \right)^{-1}\,\partial_{\sigma}X^i_{\tau'} +
    \nonumber\\
  && +\int_{\partial\Sigma} A_1(X_2) \partial_{\tau} \tilde{X}^{1}
     + i \int_{\partial\Sigma} \Phi_i(X_2) \partial_{\sigma} X^i  +
     \int_{\partial\Sigma} J_0 G^{-1}_{\tau,\tau'} X^2 + J_2
     G^{-1}_{\tau,\tau'} X^0 +
     \nonumber\\
  &&  +\int_{\partial\Sigma} J_1 G^{-1}_{\tau,\tau'} X^1 +
      J_i \left(\partial_{\tau\tau'}G_{\tau,\tau'}\right)^{-1}
      \partial_{\sigma}X^i,
\label{eq:Stot}
\eea
where $G_{\tau,\tau'}= \ln|\tau-\tau'|$. For the fields $X^i$ the boundary action is in terms of the
normal derivative $\partial_{\sigma} X^i$, since those are the boundary data for a field obeying
Dirichlet boundary conditions. As in the case of plane waves \cite{Amati:1989},
the idea is that we can integrate
in $\tilde{X}^0$ since it appears linearly in the action. This fixes $X^2$ to its classical value
$J_2$. Integrating in $X^2$ amounts to replacing $X_2\rightarrow J_2$. Afterwards all the integrals
are Gaussian and the interactions are linear in the fields, corresponding to shifting the sources
and not introducing any divergences. The $\beta$ functions will then vanish as we wanted to show.
We can also compute string diagrams as in the free theory
but we should include the appropriate non-commutative factors. A similar analysis can be done for
the superstring with the same result.

\section{Discussion}
\label{sec:conclusions}

In this paper, we have considered several different aspects of the
physics of supertubes. First, we have examined D2 supertubes
ending on an orthogonal D4-brane, by an explicit construction of
the appropriate field configuration in the worldvolume theory of
the D4-brane. {}From the latter point of view, the supertube appears
to be a dyonic string. It would be interesting to consider these
objects further as a probe of the gauge theory. For example, in
the large N limit, hanging a supertube in the throat geometry of a
collection of D4-branes could tell us about the correlation
functions of the dyonic strings, using the gravity/gauge theory
correspondence \cite{usual}. It may also be interesting to study
these configurations from the point of view of the D2-brane
worldvolume theory. For a collection of coincident D2-branes, this
theory becomes non-abelian and the expansion of the supertube into
a D4-brane can be realized using non-commutative geometry,
following the constructions of \cite{fun}.

We also considered the higher dimensional generalizations of the
supertube. In section 3, a D4 supertube was constructed where
the constituent branes included D4-branes, D2-branes and fundamental
strings. Using the freedom of rotating the D2-branes, we showed that
in this case the Born-Infeld action has supersymmetric solutions describing
supertubes with non-flat worldvolume.
These solutions are singular because of the infinite
extent of the D2-branes, however, we argued that this problem would
be avoided if the construction was generalized to a compact space.

In section 4, we constructed a family of D3 configurations where
the spatial topology was $S^1\times S^1\times R$ and the two orthogonal
circles were both supported against collapse by independent angular momenta.
These solutions were shown to be stable against small fluctuations
but were not supersymmetric. In this respect, these solutions are
rather like the nonsupersymmetric configurations of \cite{Harmark:2000na}
where ellipsoidal membranes were supported by angular momenta in
orthogonal planes. While we found a broad family of solutions, it is
interesting to note that the profiles of the magnetic field and shape
on each of the circles were correlated as in eq.~\reef{correlate}. Hence
one does not seem to have the same arbitrariness as in the case of
the D2 supertube.

Finally we examined D2 supertubes from the point of view of the worldsheet
theory of open strings. One result was that the supertubes are solutions to
all orders in $\alpha'$. Hence the arbitrariness in choosing the magnetic
field and the shape of the supertube is not lifted in string theory (at least
at lowest order in $g_s$) by $\ap$ corrections beyond those captured in
the Born-Infeld action. The way in which this result was realized that
the boundary correlators vanished for the coordinate parametrizing the
cross-section of the supertube. This was because the boundary
correlators are determined by the open-string metric described in
\cite{Seiberg:1999vs}, which was modified by the background gauge
field strengths on the D2-brane. Hence, from this point of view, the
arbitrariness in the profile of the supertube is similar to that
appearing in the profile for certain exact closed-string backgrounds
representing plane gravitational waves \cite{Amati:1989}.
It is also reminiscent of the recent discussion
of supertubes given in \cite{curve}.

We might consider the open-string metric for the configurations of
section 2 describing a supertube intersecting an orthogonal D4-brane.
To simplify the discussion, consider a flat supertube with
constant density $\rho$, which is a particular case of the examples
considered in 2.3. In this case, the gauge and scalar fields
are given by:
\be
A_0= -X_9 = -\frac{\pi\rho}{r},\ \ A_x=\frac{\pi j}{r}
\ee
where we use the notation of subsection 2.3. The induced
worldvolume metric and field strength follow as:
\bea
ds^2 &=& -dx_0^2 + dx^2 + \left(1+\frac{\pi^2\rho^2}{r^4}\right) dr^2 +
r^2d\Omega_2^2
\nonumber\\
F &=& -\frac{\pi\rho}{r^2} dx_0 \wedge dr + \frac{\pi j}{r^2} dx\wedge dr
+\pi j \sin(\theta)d\theta d\phi
\label{yes}
\eea
The open-string metric is then easily computed as
\bea
dS^2 &=& -\frac{1}{f(r)} dx_0^2 + \left(1+\frac{\pi^2 j^2}{r^4 f(r)}\right) dx^2
     -2\frac{\pi^2 j \rho}{r^4 f(r)} dx_0 dx
     \nonumber\\
     &&\quad + \left(1+\frac{\pi^2 j^2}{r^4}\right) dr^2 +
r^2 \left(1+\frac{\pi^2 j^2}{r^4}\right) d\Omega_2^2 ,
\eea
where $f(r)=1+\pi^2\rho^2/r^4$. It is clear that for $r\rightarrow \infty$
this metric is just the flat D4-brane metric, and hence the boundary
correlator for the worldsheet $x$ field will take a conventional
form. However for $r\rightarrow 0$ we obtain
\be
dS^2 \simeq \left(1+\frac{j^2}{\rho^2}\right) dx^2 -2 \frac{j}{\rho} dx_0 dx + \frac{\pi^2 j^2}{r^4} dr^2
+ \frac{\pi^2 j^2}{r^2} d\Omega^2_2
\ee
which, after identifying $B=j/\rho$ and changing coordinates to
$X_9=\pi\rho/r$ agrees both
eqs. (\ref{eq:tildecoord}), (\ref{eq:opds}).
In particular, $G^{xx}\simeq0$ and so the corresponding boundary
correlator vanishes. It would be interesting to
investigate this issue further by considering how a perturbation propagates
from the D4-brane into the supertube spike.

Finally, note that supergravity solutions corresponding to supertubes were
constructed~\cite{Emparan:2001ux}
and these turned out to be related to the so-called chiral sigma models
of~\cite{Horowitz:1994rf}, as noticed in~\cite{Lunin:2001fv}.
In fact the uplifted eleven-dimensional solution of~\cite{Emparan:2001ux} is
given by the metric and three form:
\bea
ds_{11}^2 &=& U^{-2/3}\left[ -dt^2+dz^2+K(dt+dz)^2+2(dt+dz)A+dx^2\right] +
U^{1/3} d\vec{y}d\vec{y} \nonumber\\
C_{[3]} &=& U^{-1} dt\wedge dx\wedge dz - U^{-1} (dt+dz)\wedge dx\wedge A ,
\label{eq:supergrav}
\eea
where $\vec{y}$ spans $R^8$, $U(\vec{y})$ and $K(\vec{y})$ are harmonic functions and $A(\vec{y})$ is
a harmonic 1-form in $R^8$. Dimensionally reducing along $z$ one obtains the supertube
solution~\cite{Emparan:2001ux} and dimensionally reducing in $x$ the chiral sigma model (which is exact
to all orders in $\alpha'$). Here,
we wish to remark that near the supertube the dilaton diverges and one is forced to use
the 11-dimensional perspective. However since $U\rightarrow\infty$ in this limit, the size of the
$x$ circle becomes small and one can dimensionally reduce in $x$ obtaining a chiral null model
as a near horizon description which in fact is just the near horizon limit of the fundamental
string solution.
One might also observe that in these supergravity solutions, the directions
tangent to the cross-section of the supertube become null in the near
horizon region. This may be related to the `null' behavior of the boundary
correlators discussed above, as this near horizon geometry should capture
the physics of the worldvolume theory.

\section*{Acknowledgments}

We are grateful to John Brodie, Mark Van Raamsdonk and Frederik Denef
for useful
discussions and comments. Research by AWP was supported in part by the
Canadian Institute for Advanced Research and the
National Sciences and Engineering Research Council (NSERC) of Canada.
MK, RCM and DJW are supported in part by NSERC of Canada and Fonds FCAR du
Qu\'ebec.  DJW is further supported by a McGill Major Fellowship and,
in addition, wishes to thank the Perimeter Institute and the
University of Waterloo Physics Department for
their ongoing hospitality. RCM thanks the Isaac Newton Institute of
Mathematical Sciences for their hospitality
in the final stages of this project.


\end{document}